\documentclass{ws-procs9x6}
\usepackage{mathrsfs,amsmath,amssymb,amsfonts}\input{Qcircuit}
\def\kk{\rangle\!\rangle}\def\bb{\langle\!\langle}
\def\map#1{{\mathcal{#1}}}\def\conv#1{{\mathscr{#1}}}\def\set#1{{\sf #1}}\def\sH{\set{H}}
\def\Bnd#1{\set{B(#1)}}\def\Unt#1{\set{U(#1)}}\def\SU#1{\mathbb{SU}(#1)}
\def\Reals{\mathbb R}\def\Cmplx{\mathbb C}
\def\<{\langle}\def\>{\rangle}\def\transp#1{{#1}^\intercal}\def\Tr{\operatorname{Tr}}

\def\n#1{|\!|#1|\!|}
\def\eg{e.~g. }\def\ie{i.~e. }\def\vec#1{{\boldsymbol #1}}
\begin{document} 

\markboth{G. M. D'Ariano and P. Perinotti}
{On the most efficient unitary transformation for programming quantum channels}

\title{On the most efficient unitary transformation for programming quantum
  channels\footnote{\uppercase{T}his work has been co-founded by the \uppercase{EC} under the
    program \uppercase{ATESIT} (Contract No. IST-2000-29681), and the 
    \uppercase{MIUR} {\em cofinanziamento}  2003.} }
\author{Giacomo Mauro D'Ariano\footnote{\uppercase{W}ork partially supported by the MURI program
    administered by the \uppercase{U.S. A}rmy \uppercase{R}esearch \uppercase{O}ffice under Grant
    No. \uppercase{DAAD19-00-1-0177}}  
 and Paolo Perinotti\footnote{\uppercase{W}ork partially supported by
    \uppercase{INFM} under project  \uppercase{PRA-2002-CLON}. }}
\address{{\em QUIT group}, INFM-CNR, Dipartimento di Fisica  ``A. Volta'', via Bassi 6, 27100
  Pavia, Italy\footnote{ \uppercase{P}art of the work has been carried out 
at the \uppercase{M}ax \uppercase{P}lanck \uppercase{I}nstutute for the \uppercase{P}hysics of
\uppercase{C}omplex \uppercase{S}ystems in \uppercase{D}resden during the {\em
  \uppercase{I}nternational \uppercase{S}chool of \uppercase{Q}uantum \uppercase{I}nformation}, \uppercase{S}eptember 2005.}}   
\maketitle 

\abstracts{We address the problem of finding the optimal joint unitary transformation on system +
  ancilla which is the most efficient in programming any desired channel on the system by changing
  the state of the ancilla. We present a solution to the problem for $\dim(\sH)=2$ for both
  system and ancilla.}


\section{Introduction}
A fundamental problem in quantum computing and, more generally, in quantum information
processing\cite{Nielsen} is to experimentally achieve any theoretically designed quantum channel
with a fixed device, being able to program the channel on the state of an ancilla. This problem is
of relevance for example in proving the equivalence of  cryptographic protocols, \eg proving the
equivalence between a multi-round and a single-round quantum bit commitment\cite{dsw}.
What makes the problem of channel programmability non trivial is that exact universal
programmability of channels is impossible, as a consequence of a no-go theorem for programmability
of unitary transformations by Nielsen and Chuang\cite{niels}. A similar situation occurs for
universal programmability of POVM's\cite{fiurasek2,our}. It is still possible to achieve
programmability probabilistically\cite{buzekproc}, or even deterministically\cite{ciravid},
though within some accuracy. Then, for the deterministic case, the problem is to determine the
most efficient programmability, namely the optimal dimension of the program-ancilla for given
accuracy. Recently, it has been shown \cite{our} that a dimension increasing polynomially with precision is
possible: however, even though this is a dramatical improvement compared to preliminary indications
of an exponential grow\cite{fiurasek1}, still it is not optimal. 

In establishing the theoretical limits to state-programmability of channels and POVM's the starting
problem is to find the joint system-ancilla unitary which achieves the best accuracy for fixed
dimension of the ancilla: this is exactly the problem that is addressed in the present paper. The
problem turned out to be hard, even for low dimension, and here we will give a solution for the
qubit case, for both system and ancilla.
\section{Statement of the problem}
We want to program the channel by a fixed device as follows
\begin{equation}\label{partrace}
\map{P}_{V,\sigma}(\rho)\doteq\Tr_2[V(\rho\otimes\sigma)V^\dag],
\end{equation}
with the system in the state $\rho$ interacting with an ancilla in the state $\sigma$ via the
unitary operator $V$ of the programmable device (the state of the ancilla is the {\em program}).
For fixed $V$ the above map can be regarded as a linear map from the convex set of the ancilla
states $\conv{A}$ to the convex set of channels for the system $\conv{C}$. We will denote by
$\conv{P}_{V,\conv{A}}$ the image of the ancilla states $\conv{A}$ under such linear map: these are
the programmable channels. According to the well known no-go theorem by Nielsen and Chuang it is
impossible to program all unitary channels on the system with a single $V$ and a finite-dimensional
ancilla, namely the image convex $\conv{P}_{V,\conv{A}}\subset\conv{C}$ is a proper subset of the
whole convex $\conv{C}$ of channels.  This opens the following problem:
\begin{itemize}
\item[]  {\bf Problem:} {\em For given dimension of the ancilla, find the unitary operators $V$ 
  that are the most efficient in programming channels, namely which minimize the largest distance
  $\varepsilon(V)$ of each channel $\map{C}\in{\conv{C}}$ from the programmable set
  $\conv{P}_{V,\conv{A}}$: } 
\begin{equation}\label{eps}
\varepsilon(V)\doteq\max_{\map{C}\in{\conv{C}}}\min_{\map{P}\in\conv{P}_{V,\conv{A}}}
\delta(\map{C},\map{P})\equiv\max_{\map{C}\in{\conv{C}}}\min_{\sigma\in\conv{A}}
\delta(\map{C},\map{P}_{V,\sigma}).
\end{equation}
\end{itemize}
\medskip
\par As a definition of distance it would be most appropriate to use the CB-norm distance
$\n{\map{C}-\map{P}}_{CB}$. However, this leads to a very hard problem. We will use instead the
following distance
\begin{equation}\label{del}
\delta(\map{C},\map{P})\doteq \sqrt{1-F(\map{C},\map{P})},
\end{equation}
where $F(\map{C},\map{P})$ denotes the Raginsky fidelity \cite{raginski}, which for unitary map
$\map{C}\equiv\map{U}=U\cdot U^\dag$ is equivalent to the channel fidelity \cite{Nielsen} 
\begin{equation}\label{ragfidel}
F(\map{U},\map{P})=\frac{1}{d^2}\sum_i|\Tr[C_i^\dag U]|^2,
\end{equation}
where $\map{C}=\sum_i C_i\cdot C_i^\dag$. Such fidelity is also related to the input-output fidelity
averaged over all pure states $\overline{F}_{io}(\map{U},\map{P})$, by the formula
$\overline{F}_{io}(\map{U},\map{P})=[1+dF(\map{U},\map{P})]/(d+1)$. Therefore, our optimal unitary
$V$ will maximize the fidelity
\begin{equation}\label{ragfidel2}
F(V)\doteq\min_{U\in\Unt{H}}F(U,V),\quad F(U,V)\doteq\max_{\sigma\in\conv{A}}
F(\map{U},\map{P}_{V,\sigma}) 
\end{equation}

\section{Reducing the problem to an operator norm}
In the following we will use the GNS representation $|\Psi\kk=(\Psi\otimes I)|I\kk$ of operators
$\Psi\in\Bnd{H}$, and denote by $\transp{X}$ the transposed with respect to the cyclic vector
$|I\kk$, \ie  $|\Psi\kk=(\Psi\otimes I)|I\kk=(I\otimes\transp{\Psi})|I\kk$, and by $X^*$ the complex
conjugated operator  $X^*\doteq(\transp{X})^\dag$, and write $|\upsilon^*\>$ for the vector such that 
$(|\upsilon\>\<\upsilon|\otimes I)|I\kk=|\upsilon\>|\upsilon^*\>$.
Upon spectralizing the unitary $V$ as follows
\begin{equation}
V=\sum_ke^{i\theta_k}|\Psi_k\kk\bb\Psi_k|,
\end{equation}
we obtain the Kraus operators for the map $\map{P}_{V,\sigma}(\rho)$
\begin{equation}
\map{P}_{V,\sigma}(\rho)=\sum_{nm}C_{nm}\rho C_{nm}^\dag,\qquad C_{nm}=
\sum_ke^{i\theta_k}\Psi_k|\upsilon_n^*\>\<\upsilon_m^*|\Psi_k^\dag\sqrt{\lambda_m}
\end{equation}
where $|\upsilon_n\>$ denotes the eigenvector of $\sigma$ corresponding to the eigenvalue
$\lambda_n$. We then obtain
\begin{equation}
\begin{split}
\sum_{nm}|\Tr[C_{nm}^\dag U]|^2=&
\sum_{kh}e^{i(\theta_k-\theta_h)}\Tr[\Psi_k^\dag U^\dag\Psi_k\transp{\sigma}\Psi_h^\dag U\Psi_h]\\
=&\Tr[\transp{\sigma} S(U,V)^\dag  S(U,V)]
\end{split}
\end{equation}
where 
\begin{equation}
S(U,V)=\sum_k e^{-i\theta_k}\Psi^\dag_kU\Psi_k\,.\label{esse}
\end{equation}
The fidelity (\ref{ragfidel2}) can then be rewritten as follows
\begin{equation}\label{avfuv}
F(U,V)=\frac1{d^2}\n{S(U,V)}^2.
\end{equation}
\section{Solution for the qubit case}
The operator $S(U,V)$ in Eq. (\ref{esse}) can be written as follows
\begin{equation}
S(U,V)=\Tr_1[(\transp{U}\otimes I)V^*]\,.
\end{equation}
Changing $V$ by local unitary operators transforms $S(U,V)$ in the following fashion
\begin{equation}
S(U,(W_1\otimes W_2)V(W_3\otimes W_4))=W_2^*S(W_1^\dag UW_3^\dag,V)W_4^*,
\end{equation}
namely the local unitaries do not change the minimum fidelity, since the unitaries on the ancilla
just imply a different program state, whereas the unitaries on the system just imply that the minimum
fidelity is achieved for a different unitary---say $W_1^\dag UW_3^\dag$ instead of $U$.

\begin{figure}[h]
\begin{center}
$$\Qcircuit @C=1em @R=.7em {  & \gate{W_1} 
& \ctrl{1} & \gate{X_{\alpha_1}} & \ctrl{1} & 
\gate{Z_{-\frac{\pi}{4}}} & \ctrl{1} & \gate{X_{-\alpha_2}} & \ctrl{1} & \gate{Z_{\frac{\pi}{4}}} &
\gate{W_3} & \qw \\ 
&\gate{W_2} & \targ & \gate{Z_{\alpha_3}} & \targ & \gate{Z_{-\frac{\pi}{4}}}  & \targ & \qw & \targ &
\gate{Z_{\frac{\pi}{4}}} & \gate{W_4} & \qw
}$$
\caption{Quantum circuit scheme for the general joint unitary operator $V$ in Eq. (\ref{parambipu}). Here we
  use the notation  $G_\phi=\exp(i\phi\sigma_G)$ with $G=X,Y,Z$.
\label{Vc}}
\end{center}
\end{figure}
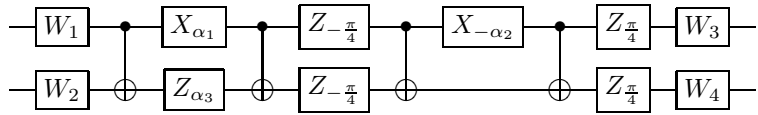

\begin{figure}[h]
\begin{center}
$$\Qcircuit @C=1em @R=.7em {  & \ctrl{1} & \gate{X_{\pm\frac{\pi}{4}}} & \ctrl{1} & \qw \\
& \targ & \gate{Z_{\mp\frac{\pi}{4}}} & \targ & \qw 
}$$
\caption{Quantum circuit scheme for the optimal unitary operator $V$ in Eq. (\ref{optV}). For the
  notation see Fig. \ref{Vc}. For the derivation of the circuit see Eqs. (\ref{circuit}).
\label{circuitV}}
\end{center}
\end{figure}
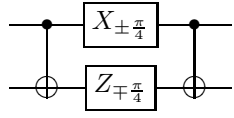

For system and ancilla both two-dimensional, one can parameterize all possible joint unitary
operators as follows\cite{KrausCirac}
\begin{equation}
V=(W_1\otimes W_2)\exp[i(\alpha_1\sigma_1\otimes\transp{\sigma_1}+\alpha_2\sigma_2\otimes\transp{\sigma_2}+\alpha_3\sigma_3\otimes\transp{\sigma_3})](W_3\otimes W_4)\,.
\label{parambipu}
\end{equation}
A possible quantum circuit to achieve $V$ in Eq. (\ref{parambipu}) can be designed using the identities
\begin{equation}\label{circuit}
\begin{split}
&[\sigma_\alpha\otimes\sigma_\alpha,\sigma_\beta\otimes\sigma_\beta]=0,\\
&C(\sigma_x\otimes I) C=\sigma_x\otimes\sigma_x,\\
&C(I\otimes \sigma_z) C=-\sigma_z\otimes\sigma_z,\\
&\left(e^{-\frac{i\pi}{4}\sigma_z}\otimes e^{-\frac{i\pi}{4}\sigma_z}\right)
 C(\sigma_x\otimes I)
 C\left(e^{\frac{i\pi}{4}\sigma_z}\otimes^{\frac{i\pi}{4}\sigma_z}\right)=\sigma_y\otimes\sigma_y,
\end{split}
\end{equation}
where $C$ denotes the controlled-NOT
\begin{equation}
C=|0\>\<0|\otimes I+|1\>\<1|\otimes\sigma_x.
\end{equation}
This gives the quantum circuit in Fig. \ref{Vc}.
The problem is now reduced to study only joint unitary operators of the form
\begin{equation}
V=\exp[(i(\alpha_1\sigma_1\otimes\transp{\sigma_1}+\alpha_2\sigma_2\otimes\transp{\sigma_2}+\alpha_3\sigma_3\otimes\transp{\sigma_3})]\,.
\end{equation}
This has eigenvectors 
\begin{equation}
|\Psi_j\kk=\frac1{\sqrt2}|\sigma_j\kk,
\end{equation}
where $\sigma_j$, $j=0,1,2,3$ denote the Pauli matrices $\sigma_0=I$, $\sigma_1=\sigma_x$,
$\sigma_2=\sigma_y$, $\sigma_3=\sigma_z$. This means that we can rewrite $S(U,V)$ in
Eq.~\eqref{esse} as follows 
\begin{equation}
S(U,V)=\frac12\sum_{j=0}^3e^{-i\theta_j}\sigma_j U\sigma_j\,,
\end{equation}
with
\begin{equation}
\theta_0=\alpha_1+\alpha_2+\alpha_3\,,\quad\theta_i=2\alpha_i-\theta_0\,.
\end{equation}
The unitary $U$ belongs to $\SU2$, and can be written in the Bloch form
\begin{equation}\label{Bloch}
U=n_0I+i\vec n\cdot\vec\sigma\,,
\end{equation}
with $n_k\in\Reals$ and $n_0^2+|\vec n|^2=1$. Using the identity
\begin{equation}
\sigma_j \sigma_l \sigma_j=\epsilon_{jl}\sigma_l,\qquad
\epsilon_{j0}=\epsilon_{jj}=1,\quad\epsilon_{jl}=-1\,, l\neq 0,j,
\end{equation}
we can rewrite
\begin{equation}
S(U,V)=\tilde n_0 I+\tilde{\vec n}\cdot\vec\sigma,
\end{equation}
where 
\begin{equation}
\begin{split}
\tilde n_j=&t_jn_j,\quad 0\leq j\leq 3, t_0=\frac12\sum_{j=0}^3e^{-i\theta_j},\\ 
t_j=&e^{-i\theta_0}+e^{-i\theta_j}-t_0,\;1\leq j\leq 3,\qquad t_j=|t_j|e^{i\phi_j},\; 0\leq j\leq 3, \\ 
\end{split}
\end{equation}
It is now easy to evaluate the operator $S(U,V)^\dag S(U,V)$. One has
\begin{equation}
\begin{split}
S(U,V)^\dag S(U,V)=&v_0 I+ \vec v\cdot\vec \sigma,\\
v_0=&|\tilde n_0|^2+|\tilde{\vec n}|^2,\quad
\vec v=i\left[2\Im(\tilde n_0\tilde{\vec n}^*)+\tilde{\vec n}^*\times\tilde{\vec n}\right]\,.
\end{split}
\end{equation}
Now, the maximum eigenvalue of $S(U,V)^\dag S(U,V)$ is $v_0+|\vec v|$, and one has
\begin{equation}
|\vec v|^2=\sum_{i,j=0}^3|\tilde n_i|^2|\tilde n_j|^2-\tilde n_i^{*2}\tilde
n_j^2=2\sum_{i,j=0}^3|\tilde n_i|^2|\tilde n_j|^2\sin^2(\phi_i-\phi_j),
\end{equation}
whence the norm of $S(U,V)$ is given by
\begin{equation}\label{SUV}
\n{S(U,V)}^2=\sum_{j=0}^3n_j^2|t_j|^2+\sqrt{2\sum_{i,j=0}^3n_i^2n_j^2|t_i|^2|t_j|^2\sin^2(\phi_i-\phi_j)}\,.
\end{equation}
Notice that the unitary $U$ which is programmed with minimum fidelity in general will not not be
unique, since the expression for the fidelity depends on $\{n_j^2\}$. Notice also that using the
decomposition in Eq.~\eqref{parambipu} the minimum fidelity just depends on the phases
$\{\theta_j\}$, and the local unitaries will appear only in the definitions of the optimal program
state and of the worstly approximated unitary.  It is convenient to write Eq. (\ref{SUV}) as follows
\begin{equation}\label{SUV2}
\n{S(U,V)}^2=\vec u\cdot\vec t+\sqrt{ \vec u\cdot\vec T\vec u}\,.
\end{equation}
where $\vec u=(n_0^2 ,n_1^2 ,n_2^2 ,n_3^2 )$, $\vec t=(|t_0|^2 ,|t_1|^2 ,|t_2|^2 ,|t_3|^2)$, and
$\vec T_{ij}=|t_i|^2|t_j|^2\sin^2(\phi_i-\phi_j)$. One has the bounds
\begin{equation}
\vec u\cdot\vec t+\sqrt{ \vec u\cdot\vec T\vec u}\geq \vec u\cdot\vec t\geq \min_j |t_j|^2, 
\end{equation}
and the bound is achieved on one of the for extremal points $u_l=\delta_{lj}$ of the domain of $\vec
u$ which is the convex set $\{\vec u,\; u_j\geq 0,\,\sum_j u_j=1\}$ (the positive octant of the unit
four dimensional ball $S^4_+$). Therefore, the fidelity minimized over all unitaries is given by
\begin{equation}
F(V)=\frac1{d^2}\min_j|t_j|^2.
\end{equation}
The optimal unitary $V$ is now obtained by maximizing $F(V)$. We need then to consider the
decomposition Eq.~\eqref{parambipu}, and then to maximize the minimum among the four eigenvalues of
$S(U,V)^\dag S(U,V)$.  Notice that $t_j=\sum_{\mu}H_{j\mu}e^{i\theta_\mu}$, where $H$ is the
Hadamard matrix
\begin{equation}
H=\frac12
\begin{pmatrix}
1&1&1&1\\
1&1&-1&-1\\
1&-1&1&-1\\
1&-1&-1&1
\end{pmatrix},
\end{equation}
which is unitary, and consequently
$\sum_j|t_j|^2=\sum_j |e^{i\theta_j}|^2=4$.
This implies that $\min_j|t_j|\leq1$. We now provide a
choice of phases $\theta_j$ such that $|t_j|=1$ for all
$j$, achieving the maximum fidelity allowed. For instance, we
can take $\theta_0=0,\theta_1=\pi/2,\theta_2=\pi,\theta_3=\pi/2$, corresponding to
the eigenvalues $i,1,-i,1$ for $V$. Another solution is
$\theta_0=0,\theta_1=-\pi/2,\theta_2=\pi,\theta_3=-\pi/2$. Also one can set $\theta_i\to
-\theta_i$. The eigenvalues of $S(U,V)^\dag S(U,V)$ are then $1,1,1,1$, while for the fidelity we have 
\begin{equation}\label{optV}
F\doteq\max_{V\in\Unt{H^{\otimes 2}}}F(V)=\frac{1}{d^2}=\frac{1}{4},
\end{equation}
and the corresponding optimal $V$ has the form
\begin{equation}
V=\exp\left[\pm i\frac{\pi}{4}\left(\sigma_x\otimes\sigma_x\pm\sigma_z\otimes\sigma_z\right)\right].
\end{equation}
A possible circuit scheme for the optimal $V$ is given in Fig. \ref{circuitV}.

\par We now show that such fidelity cannot be achieved by any $V$ of the controlled-unitary form
\begin{equation}
V=\sum_{k=1}^2V_k\otimes|\psi_k\>\<\psi_k|,\qquad \<\psi_1|\psi_2\>=0,\quad V_1, V_2\hbox{ unitary
  on }\sH\simeq\Cmplx^2.
\end{equation}
For spectral decomposition $V_k=\sum_{j=1}^2e^{i\theta^{(j)}_k}|\phi^{(k)}_j\>\<\phi^{(k)}_j|$ the
eigenvectors of $V$ are  $|\Psi_{jk}\kk=|\phi^{(k)}_j\>|\psi_k\>$,  and the corresponding operators
are $\Psi_{jk}=|\phi^{(k)}_j\>\<\psi_k^*|$, namely the operator $S(U,V)$ is
\begin{equation}
S(U,V)=\sum_{j,k}e^{-i\theta^{(j)}_k}|\psi_k^*\>\<\phi^{(k)}_j|U|\phi^{(k)}_j\>\<\psi_k^*|\,,
\end{equation}
with singular values $\sum_{j=1}^2e^{-i\theta^{(j)}_k}\<\phi^{(k)}_j|U|\phi^{(k)}_j\>=\Tr[V_k^\dag U]$.
Then, the optimal program state is $|\psi_h\>$, with $h=\arg\max_k|\Tr[V_k^\dag U]|$, and the
corresponding fidelity is 
\begin{equation}
F(U,V)=\frac{1}{4}|\Tr[V_h^\dag U]|^2\,,
\end{equation}
and one has
\begin{equation}
F(V)=\min_U F(U,V)=0,
\end{equation}
since for any couple of unitaries $V_k$ there always exists a unitary $U$ such that $\Tr[V^\dag_k
U]=0$ for $k=1,2$. Indeed, writing the unitaries in the Bloch form (\ref{Bloch}), their
Hilbert-Schmidt scalar is equal to the euclidean scalar product in $\Reals^4$ of their corresponding
vectors, whence it is always possible to find a vector orthogonal to any given couple in
$\Reals^4$. The corresponding $U$ is then orthogonal to both $V_k$, and the minimum fidelity for any
controlled-unitary is zero.

\end{document}